\documentclass{moriond}

\def\Journal#1#2#3#4{{#1} {\bf #2}, #3 (#4)}

\def\NIMA{{\em Nucl. Instrum. Methods} A}

\def\PRL{\em Phys. Rev. Lett.}
\def\PRD{{\em Phys. Rev.} D}

\def\be{\begin{equation}}
\def\ee{\end{equation}}
\def\bea{\begin{eqnarray}}
\def\eea{\end{eqnarray}}

\newcommand{\Photo}{}
\usepackage{amsmath}
\usepackage{amssymb}

\begin{document}
\vspace*{4cm}
\title{Results and Prospects of T2K}

\author{T. Nosek (on behalf of the T2K collaboration)}

\address{High Energy Physics Division, National Centre for Nuclear Research,\\
Ludwika Pasteura 7, 02-093 Warsaw, Poland}

\maketitle\abstracts{T2K is a neutrino oscillation experiment with a 295 km long baseline between the far detector, Super-Kamiokande, and a suite of near detectors to study $\nu_\mu/\bar{\nu}_\mu$ disappearance and $\nu_e/\bar{\nu}_e$ appearance in a $\nu_\mu/\bar{\nu}_\mu$ neutrino beam produced at J-PARC. The experiment has excluded CP conservation in the three-neutrino oscillation model at $>90~\%$ CL and precisely constrained the parameters $\vert \Delta m_{32}^2 \vert$ and $\sin^2 \theta_{23}$. This paper reports on the novelties of the most up-to-date 2022 analysis and the near-future prospects of T2K.}

\section{Introduction}

As of today, the observed neutrino oscillation phenomena can be well described within the mixing model of three massive neutrinos in three active flavor eigenstates of the Standard Model: electron ($\nu_e$), muon ($\nu_\mu$), and tau ($\nu_\tau$) neutrinos ($\nu$). This 3$\nu$-model is represented by two independent squared mass-splittings $\Delta m_{21}^2, \Delta m_{32}^2$ of the neutrino masses and a conveniently parameterized PMNS mixing matrix with three mixing angles $\theta_{12}, \theta_{13}, \theta_{23}$ and a charge-parity (CP) violating phase $\delta_\mathrm{CP}$.\cite{3nus}

The Tokai-to-Kamioka (T2K) experiment is a long-baseline neutrino oscillation experiment studying $\nu/\bar{\nu}$ flavor transitions of $\nu_\mu \to \nu_\mu$ and $\nu_\mu \to \nu_e$ in both $\nu_\mu$ and $\bar{\nu}_\mu$ beams. Its main goal is to measure $\theta_{23}$, $\Delta m_{32}^2$, and $\delta_\mathrm{CP}$, thus to target the important remaining questions of the 3$\nu$-oscillation model about the potential CP violation in the leptonic sector (whether $\delta_\mathrm{CP} \neq 0, \pi$), the precise value of $\theta_{23}$ ($\lesseqgtr 45^\circ$), and the sign of $\Delta m_{32}^2$ (normal ordering, $\Delta m_{32}^2 > 0$, or inverted ordering, $\Delta m_{32}^2 < 0$, of the neutrino masses).

This paper reports on the 2022 round of the T2K neutrino oscillation analysis, its novelties, and updates. It also presents T2K's near-future plans, ongoing upgrades, and inter-collaborative efforts, all to fully exploit its potential within the subject matter. 

\section{The T2K Experiment}

T2K uses a conventional beam of neutrinos that are born from decays of mesons (mostly $\pi$ and $K$) produced in collisions of accelerated 30 GeV protons with a graphite target at J-PARC (Japan Proton Accelerator Research Center).\cite{theT2K} The mesons are focused by three magnetic horns capable of switching polarity to select either positively or negatively charged particles and thus create a $\nu_\mu$ or $\bar{\nu}_\mu$ dominated beam, respectively.

The 50 kt water Cherenkov far detector, well-known Super-Kamiokande (SK),\cite{theSK} is located at a distance of 295~km from the J-PARC neutrino source, shifted $2.5^\circ$ off the beam axis to obtain a narrow $\nu$ energy spectra peaked around 650 MeV. This combination corresponds to the first $\nu_\mu \to \nu_\mu$ disappearance minimum and $\nu_\mu \to \nu_e$ appearance maximum, which makes T2K sensitive to $\sin^2 \theta_{23}, \sin^2 \theta_{13}, \vert \Delta m_{32}^2\vert$, and $\delta_\mathrm{CP}$ parameters. SK detects rings of Cherenkov light from superluminal $\mu/e$ emerging from charged-current (CC) neutrino interactions. From the collected hit distributions of the rings, these can be distinguished as $\mu/e$-like with a very low misidentification rate of $<0.1 \%$. By assuming quasi-elastic (QE) scattering, neutrino energy is reconstructed from the charged lepton direction and momentum.

To further characterize the beam and to measure neutrino interaction rates, T2K employs a suite of near detectors. First, INGRID (Interactive Neutrino GRID)  is an iron-scintillator sandwich detector to monitor the beam intensity in on-axis direction. Second, ND280, similarly to SK, sits $2.5^\circ$ off-axis to directly constrain neutrino fluxes and cross sections in the oscillation analysis. It is a magnetized and versatile assortment of multiple sub-detectors, including water (primarily oxygen) and active scintillator (primarily carbon) segmented targets (Fine-Grained Detectors, FGD), and time projection chambers (TPC) for tracking. Altogether, ND280 allows for reconstructing the momentum, sign, and energy loss of charged particles.

\section{2022 Neutrino Oscillation Analysis}

The 2022 T2K neutrino oscillation analysis is a follow-up on its previous round of 2020,\cite{OA2020} sharing the same SK exposure of $19.7/16.3 \times 10^{20}$ protons-on-target (POT) in $\nu/\bar{\nu}$-beam. Similarly, to enhance robustness, the analysis is carried out twice based on two methods of statistical inference. The frequentist approach proceeds in two major steps. First, it estimates the flux and neutrino cross sections from ND280 data; then, the oscillation parameters are extracted while generating SK predictions from this constrained model. The Bayesian approach employing Markov Chain Monte Carlo with the Metropolis-Hastings algorithm~\cite{MCMC} can, in contrast, constrain all the parameters in all the ND280 and SK data samples at once. For more details and general context, please, consult the 2020 paper.\cite{OA2020}

\subsection{Simulation Updates}

The base neutrino flux simulation utilizing FLUKA and GEANT3 was updated and reweighted to the hadron-production measurements from NA61/SHINE with the T2K replica target.\cite{NA61} Superseding the previously used~\cite{OA2020} NA61 dataset~\cite{NA61old} by adding more statistics and $K$ production data allowed for a modest overall decrease of flux systematic uncertainties.

When updating the NEUT neutrino interaction model, the emphasis was placed on replacing the existing heuristic systematic uncertainty parametrizations with better-motivated theory-driven ones. The treatment of the CC QE spectral function was improved to include normalization for individual nuclear shells, and Pauli blocking to allow for more freedom in the region of low four-momentum transfer.\cite{spectfunc} Also, additional uncertainties for resonant scattering (RES), multi-$\pi$ events and final state interactions were implemented.

\subsection{ND280 Samples and the New Tagging of Protons and Photons}

The basic motivation behind the ND280 samples is to determine the multiplicity of $\pi$ in the CC interaction final states, as in 2020~\cite{OA2020} -- no $\pi$ (CC0$\pi$) dominated by CC QE, one $\pi$ (CC1$\pi$) dominated by RES, and any other set of particles (CCOther), which were further broken down by the $\nu/\bar{\nu}$-beam, the FGD target of pure scintillator or scintillator interlaid with water, and the lepton charge. The 2022 analysis introduces new $\nu$-beam samples with tagged protons (CC0$\pi$+$Np$) and photons (CC$\gamma$). As $\gamma$ comes mostly from $\pi^0$ decays and deep inelastic scattering events, filtering them out enhances the purity of the remaining samples by about 10~\%. Next, the CC0$\pi$+$Np$ sample probes different regions of the energy--momentum transfer plane than CC0$\pi$+$0p$ with no tagged protons (or the original catch-all CC0$\pi$), it has lower $\pi^+$ background, and it better constrains intrinsic nuclear effects in the model.

\subsection{SK Samples and the New Multi-Ring Sample}

SK selection samples exploit the reconstructed Cherenkov rings and their $\mu/e$ PID distinguishing the flavor of the secondary charged lepton from the $\nu/\bar{\nu}$ CC interactions. There are four so-called one-ring (1R) samples, two in $\nu$-beam for $\nu_\mu$-like (1R$_\mu$) and $\nu_e$-like (1R$_e$) selection and two in $\bar{\nu}$-beam. Furthermore, one extra $\nu$-beam 1R$_e$d$_e$ sample selects delayed hit clusters consistent with Michel $e$ (d$_e$) from unseen $\pi\to\mu\to e$ decays in $\nu_e$ events.

\begin{figure}[t]
    \centering
    \includegraphics[width=.515\linewidth]{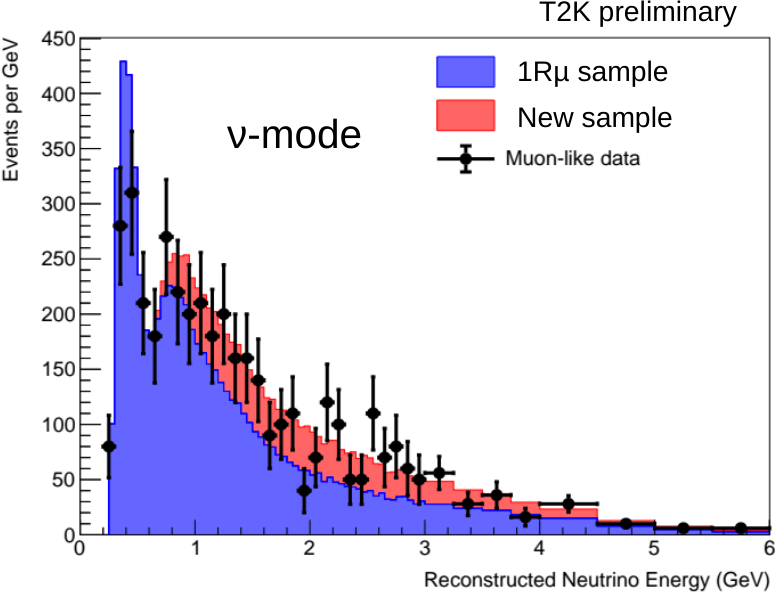}
    \includegraphics[width=.465\linewidth]{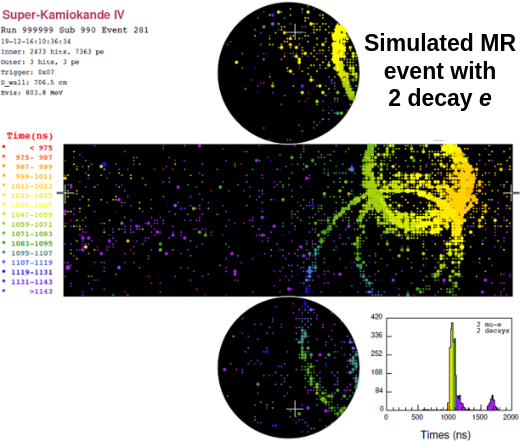}
    \caption{2022 best-fit predictions of the reconstructed neutrino energy for 1R$_\mu$ (blue) and MR$_\mu$ (red, ``New sample'') samples overlaid with the combined $\nu$-beam $\nu_\mu$-like data (left). SK event display of a simulated MR$_\mu$ with $\pi^+\to\mu^+$ both exceeding the Cherenkov threshold and with two tagged Michel $e$ from both $\mu^-$ and $\mu^+$ decays (right).}
    \label{fig:multiring}
\end{figure}

2022 analysis offers one additional $\nu$-beam sample incorporating SK multi-ring events (MR$_\mu$) to tag topologies of
\begin{itemize}
    \item one $\mu$ ring, one $\pi$ ring, and one or two Michel $e$, or
    \item one $\mu$ ring and two Michel $e$.
\end{itemize}
This MR$_\mu$ aims for $\nu_\mu$ CC events with $\pi^+$ in the final state, and it is estimated to add more than 30~\% statistics to the overall $\nu_\mu$-like $\nu$-beam selection data. Though the sample is sensitive to neutrino oscillation, with typical reconstructed energy in the tail of the spectrum above the oscillation minimum, it also provides additional modelling cross-checks and constrains backgrounds. Fig.~\ref{fig:multiring} shows a combined 1R$_\mu$+MR$_\mu$ sample prediction with the selected data and a simulated MR$_\mu$ SK event.

\subsection{Results}
\begin{figure}
    \centering
    \includegraphics[width=.46\linewidth]{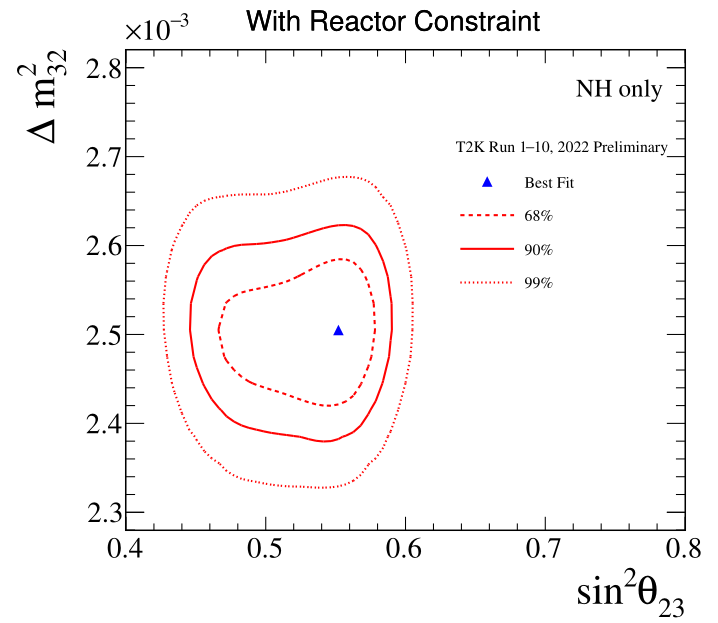}
    \includegraphics[width=.52\linewidth]{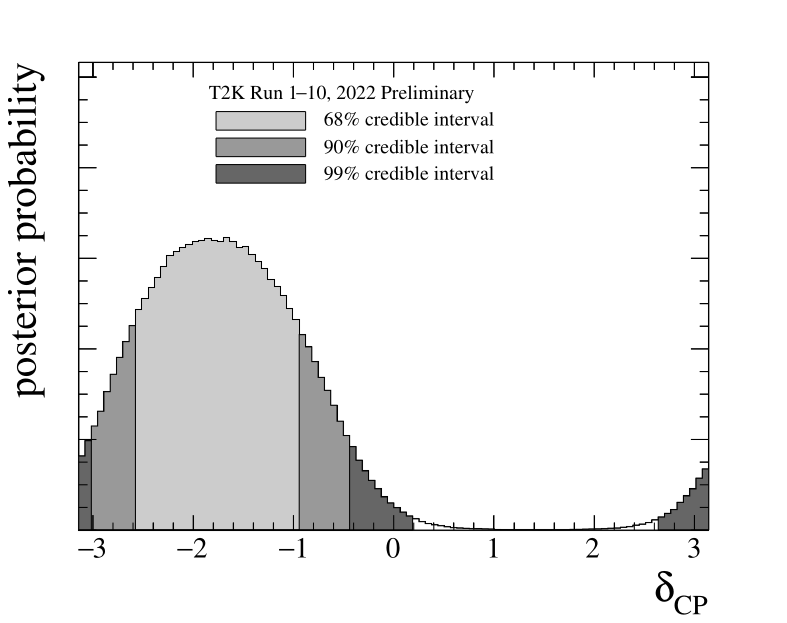}
    \caption{2022 best-fit values (blue triangle) and credible regions (red lines) in the $\Delta m_{32}^2$--$\sin^2 \theta_{23}$ upper-plane of the 3$\nu$-oscillation parameter space corresponding to the normal ordering (hierarchy, NH, $\Delta m_{32}^2 >0$) of the neutrino masses (left). Bayesian posterior probability for $\delta_\mathrm{CP}$ parameter with 68~\% (light gray), 90~\% (gray), and 99~\% (dark gray) credible intervals highlighted (right).}
    \label{fig:results}
\end{figure}

Both frequentist and Bayesian analyses use $\Delta m_{21}^2$ and $\sin^2 \theta_{12}$ constraints from the report of the Particle Data Group,\cite{PDG2021} and also $\sin^2 \theta_{13}$ constraints from the reactor neutrino experiments when delivering the ultimate statements on $\delta_\mathrm{CP}$ and $\sin^2 \theta_{23}$. Fig.~\ref{fig:results} presents the credible regions in the $\Delta m_{32}^2$--$\sin^2 \theta_{23}$ plane and the $\delta_\mathrm{CP}$ posterior probability obtained with the Bayesian MCMC.

There is a weak preference for the normal ordering of neutrino masses (Bayes factor 3.46) and also for $\sin^2 \theta_{23} > 0.5$ (Bayes factor 2.51). The data excludes CP conserving $\delta_\mathrm{CP} = 0,\pi$ at $> 90~\%$ confidence level, and both values are outside of the Bayesian 90~\% credible intervals. We note that the frequentist and Bayesian results are in excellent agreement with each other as well as with the previous analysis.\cite{OA2020} The oscillation parameters' credible regions are generally consistent with the measurements of the related neutrino oscillation experiments.\cite{NOvA2020,SKatm,MINOS,IceCube} The combined T2K model of detectors, neutrino fluxes, interactions, and oscillation within the 3$\nu$-oscillation paradigm reasonably concurs with the observed SK data as indicated by frequentist $p$-value 0.35 and Bayesian posterior predictive $p$-value 0.85.

\section{Future Prospects}

Before the dawn of the next-generation neutrino oscillation experiments scheduled in the late 2020s, T2K has planned several upgrades and joint analyses with currently active collaborations to maintain valuable progress in the subject fields of neutrino physics.

\subsection{Joint Analyses with NOvA and SK}

\begin{figure}[t]
    \centering
    \includegraphics[width=\linewidth]{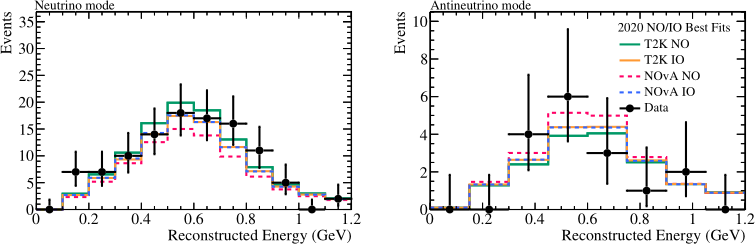}
    \caption{T2K SK 1R$_e$(1d$_e$) $\nu_e$-like selected data (black) in $\nu$-beam (left) and $\bar{\nu}$-beam (right) overlayed with the predictions of T2K 2020~$^\mathrm{4}$ (solid) and NOvA 2020~$^\mathrm{10}$ (dashed) best-fit oscillation parameters values corresponding to the normal (teal and red) and inverted (orange and blue) orderings of the neutrino masses.}
    \label{fig:t2knova}
\end{figure}

There are two important joint analyses within the 3$\nu$-oscillation model approaching their late stages of development. The NOvA+T2K analysis, based on the 2020 results,\cite{OA2020,NOvA2020}  brings together the two running long-baseline accelerator neutrino experiments. The combination of similar physics concepts, distinct detector technologies, interaction models, complementary experimental setups of different oscillation baselines (NOvA 810 km, T2K 295 km) and energies (NOvA $\sim$2 GeV, T2K $\sim$650 MeV) promises to improve the $\delta_\mathrm{CP}$ and $\sin^2 \theta_{23}$ sensitivities significantly, and the parameters constraints might further tighten eventually. Together, the NOvA+T2K data has the potential to break the mass ordering and $\theta_{23}$ octant degeneracies, otherwise implicitly present in either experiment alone. Fig.~\ref{fig:t2knova} illustrates how predictions of best-fit oscillation parameters values from T2K and NOvA 2020 analyses adhere to each other and to the T2K $\nu_e$-like selected data under the normal or inverted neutrino mass ordering hypothesis.

The second joint fit combines the T2K beam and the SK atmospheric data.\cite{SKatm} The atmospheric neutrino samples cover much wider ranges of oscillation baselines and energies, making them better sensitive to mass ordering. The conceivable weakening of this degeneracy would, in turn, enhance the $\delta_\mathrm{CP}$ sensitivity over the whole [$-\pi,+\pi$] interval.

\subsection{Beam and ND280 Upgrade}

To achieve even higher significance when it comes to the measurement of extreme values of $\delta_\mathrm{CP}$ and excluding CP conservation in the leptonic sector, T2K launched an extensive upgrade program. To increase the data statistics, this would include a J-PARC beam upgrade to reach higher powers of operation, going from the current $\sim$500 kW to more than 750~kW by the next year and later over 1 MW, eventually achieving 1.3 MW. The projected accumulated exposure should surpass 2$\times 10^{22}$~POT in $\nu+\bar{\nu}$-beam combined by 2028.\cite{beamUP}

Larger statistics require even better control of the systematic uncertainties; hence the upgrade also includes substantial modifications to ND280.\cite{ND280UP} Its upstream part, initially devoted to the measurements of background neutral current neutrino interactions with a single $\pi^0$ in their final state, will be replaced by a set of new advanced detectors. It consists of a plastic scintillator Super-FGD, segmented in more than 2 million 1 cm$^3$ cubes used as a new target, which will be sandwiched by two horizontally aligned TPCs to track high-angle $\mu$ (so-called High-Angle TPCs). From each side, all shall be covered by six time-of-flight detectors (TOF) to trigger cosmic and other entering backgrounds.

\begin{figure}[t]
    \centering
    \includegraphics[width=.48\linewidth]{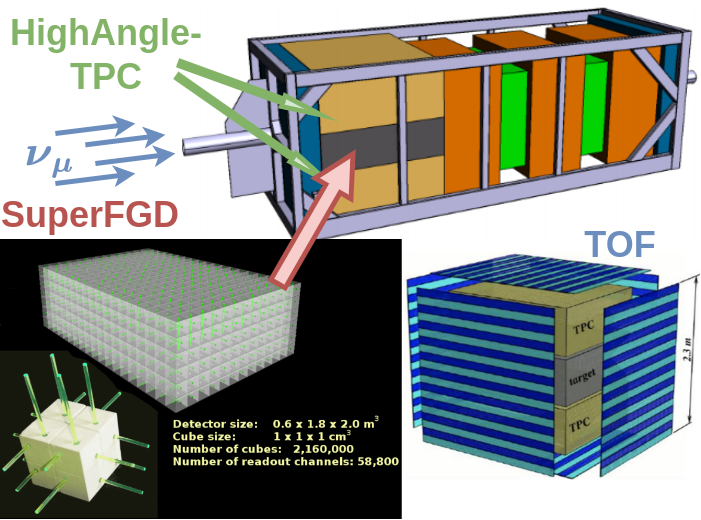}
    \includegraphics[width=.51\linewidth]{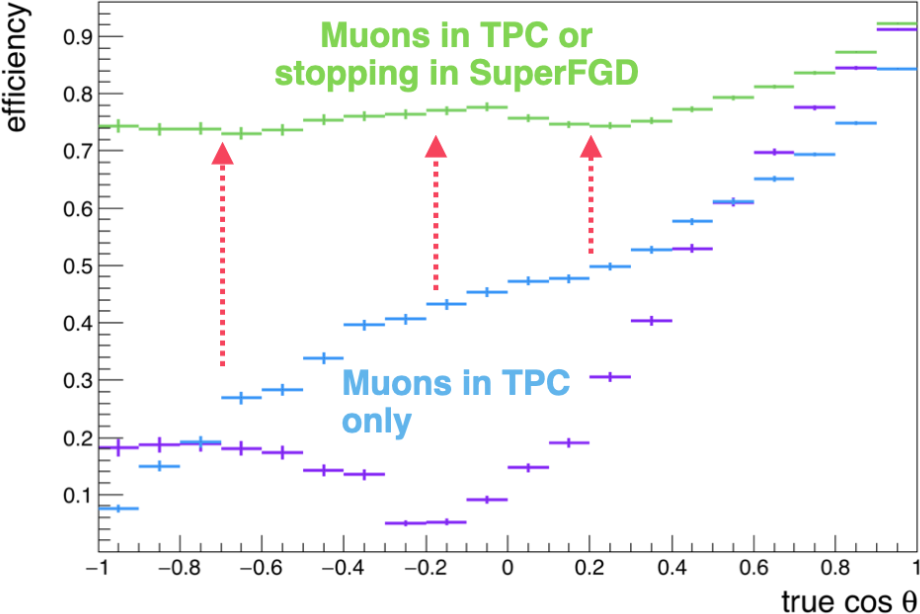}
    \caption{Sketch of the upgraded ND280 (left) -- the upstream part of ND280 will consist of a new highly-segmented scintillator target Super-FGD sandwiched between two horizontal TPCs, or High-Angle TPCs, all covered by six planes of TOF detectors. This will allow for the enhanced acceptance of $\mu$ from $\nu/\bar{\nu}$ interactions in terms of their selection efficiency as a function of the angle with respect to the event's longitudinal axis (in the plot on the right) when detected by TPCs (blue), or even stopping in the Super-FGD (green), in comparison to the regular T2K analysis of the ND280 data (purple). }
    \label{fig:nd280upgrade}
\end{figure}

The upgrade primarily enhances the ND280 angular acceptance for $\mu$ in neutrino CC events, as depicted in Fig.~\ref{fig:nd280upgrade}, and it also lowers hadronic thresholds thanks to the Super-FGD (super-fine) granularity. This allows for covering new regions of the interaction phase space and provides an advantage of using transverse kinematic hadron variables (like missing transverse momentum or visible energy) for reconstruction. Complementing the usual T2K analysis of ND280 data, this is expected to result in increased sensitivity to nuclear-model uncertainties and their eventual reduction.\cite{NDupSens}

\section{Conclusion}

The latest 2022 T2K neutrino oscillation analysis benefits from the simulation updates, new near detector samples to better constrain neutrino cross sections and fluxes, and an additional SK multi-ring sample. The results are consistent with previous T2K analyses, as well as other related oscillation experiments. The CP conserving values of $\delta_\mathrm{CP}$ are disfavored at more than 90~\% confidence level; the data very slightly prefer $\theta>45^\circ$ and the normal ordering of the neutrino masses.

Two major analyses with NOvA and SK atmospheric data are expected to further deepen T2K's direct impact and advancement in the subject. The upgrades of the T2K beam and ND280 are fully underway to increase statistics and to provide even more detailed measurements of neutrino interactions to help in reducing associated systematic uncertainties.

\section*{Acknowledgments}
This work was supported by the National Science Centre (UMO-2018/30/E/ST2/00441), Poland.


\section*{References}

\begin{thebibliography}{99}
\bibitem{3nus} C. Giunti, M. Laveder, e-Print \href{https://doi.org/10.48550/arXiv.hep-ph/0310238}{arXiv:hep-ph/0310238 [hep-ph] (2003)}.

\bibitem{theT2K} K. Abe {\it et al} [T2K], \href{https://doi.org/10.1016/j.nima.2011.06.067}{\Journal{\NIMA}{659}{106–135}{2011}}.

\bibitem{theSK} Y. Fukuda {\it et al} [Super-Kamiokande], \href{https://doi.org/10.1016/S0168-9002(03)00425-X}{\Journal{\NIMA}{501}{418-462}{2003}}.

\bibitem{OA2020} K. Abe {\it et al} [T2K], e-Print \href{https://arxiv.org/abs/2303.03222}{arXiv:2303.03222 [hep-ex] (2023)}

\bibitem{MCMC} W. K. Hastings, \href{https://doi.org/10.1093/biomet/57.1.97}{{\em Biometrika} \textbf{57} no. 1, 97–109 (1970)}.

\bibitem{PDG2021} P. A. Zyla {\it et al} [Particle Data Group], \href{https://pdg.lbl.gov/2021/}{{\em PTEP} \textbf{2020}, 083C01 (2020)}.

\bibitem{NA61} N. Abgrall {\it et al} [NA61/SHINE], \href{https://doi.org/10.1140/epjc/s10052-019-6583-0}{\Journal{{\em Eur. Phys. J.} C}{79}{100}{2019}}.

\bibitem{NA61old} N. Abgrall {\it et al} [NA61/SHINE], \href{https://doi.org/10.1140/epjc/s10052-016-4440-y}{\Journal{{\em Eur. Phys. J.} C}{76}{617}{2016}}.

\bibitem{spectfunc} O. Benhar, D. Meloni, \href{https://doi.org/10.1016/j.nuclphysa.2007.02.015}{\Journal{{\em Nucl. Phys.} A}{789}{379}{2007}}.

\bibitem{NOvA2020} M. A. Acero {\it et al} [NOvA], \href{https://doi.org/10.1103/PhysRevD.106.032004}{\Journal{\PRD}{106}{032004}{2022}}.

\bibitem{SKatm} M. Jiang {\it et al} [Super-Kamiokande], \href{https://doi.org/10.1093/ptep/ptz015}{{\em PTEP} \textbf{2019} vol. 5, 053F01 (2019)}.

\bibitem{MINOS} P. Adamson {\it et al} [MINOS+], \href{https://doi.org/10.1103/PhysRevLett.125.131802}{\Journal{\PRL}{125}{131802}{2020}}.

\bibitem{IceCube} M. G. Aartsen {\it et al} [IceCube], \href{https://doi.org/10.1103/PhysRevLett.120.071801}{\Journal{\PRL}{120}{071801}{2018}}.

\bibitem{beamUP} K. Abe {\it et al}, e-Print \href{https://doi.org/10.48550/arXiv.1908.05141}{arXiv:1908.05141 [physics.ins-det] (2019)}.

\bibitem{ND280UP} K. Abe {\it et al}, e-Print \href{https://doi.org/10.48550/arXiv.1901.03750}{arXiv:1901.03750 [physics.ins-det] (2020)}.

\bibitem{NDupSens} S. Dolan {\it et al}, \href{https://doi.org/10.1103/PhysRevD.105.032010}{\Journal{\PRD}{105}{032010}{2022}}.

\end{thebibliography}
\end{document}